\documentclass[12pt,epsf]{article}

\begin{document}

\title{Nuclear Disintegration in Magnetic Fields}
\author{Lee C. Loveridge \\ {\small \it Department of Physics and Astronomy} \\ 
			{\small \it UCLA, Los Angeles, CA 90095-1547}}
\date {UCLA/02/TEP/15 \\ \today}
\maketitle

\begin{abstract}
We employ the Weizs\"{a}cker-Williams method of virtual quanta to study
disintegration of nuclei in magnetic field.  We explore a variety of field
configurations and conclude that for the energy range of interest for
applications to cosmic rays ($10^{18}-10^{21}$ eV) such disintegrations are
not a significant source of energy or flux loss for any realistic
acceleration mechanism.
\end{abstract}

\section {Introduction}

Understanding the interactions of cosmic rays with cosmic backgrounds is
essential for solving the puzzle of ultrahigh-energy cosmic
rays~\cite{puzzle}, as well as for understanding the most powerful natural
accelerators, such as active galactic nuclei, gamma-ray bursts, {\em etc.}
Since their discovery, cosmic rays with energies beyond the
Greisen-Zatsepin-Kuzmin cutoff~\cite{gzk}, $E>10^{19.6}$eV,  have presented 
astrophysics with several important and unresolved problems.  
The composition of ultrahigh-energy cosmic rays remains unknown.  In
particular, they may be protons, photons, or nuclei.  In the latter case,
interactions with the background photons cause nuclear disintagration 
~\cite{cross-sections}.  

A number of possible acceleration mechanisms have been proposed.  Some of 
them, such as neutron stars \cite{neutronstar},
 have very large magnetic fields.  To a highly 
relativistic particle, such magnetic fields are much larger and they are 
accompanied by a matching electric field.  Thus, the magnetic field acts like
an intense photon field.  It is therefore natural to ask whether such magnetic
fields may cause energy losses in relativistic nuclei, through 
photodisintegration.    

In this paper we explore the possibility that the static magnetic fields of
an accelerator may cause photodisintegration of relativistic nuclei.
We will use the Weizs\"{a}cker-Williams method of virtual quanta, as
described in reference~\cite{Jackson}.  The same method has been used 
to study photodisintegration of nuclei in the Coulomb elctric fields
of target nuclei~\cite{Yodh}.  In sections 2-4 we set up the method by first
transforming the magnetic field to the rest frame of the accelerated
particle, second, decomposing this field as a collection of virtual
photons, and third, studying the effect of this spectrum of virtual photons
on the propagating nuclei.  In section 5 we use our method to find the
effective photon spectrum of a variety of magnetic fields.  In section 6 we
use some simple cross section data to place limits on the required magnetic
fields to get significant photodisintegration.  In section 7 we generalize
the method to apply to accelerating particles.  Finally in section 8 we
conclude.

\section {Transformation of the Magnetic Field}

If we attempt to study the disintegration of a relativistic nucleus in a 
static magnetic field from the rest frame of the field, we run into a number
of problems.  In this frame the field does not look like a collection of 
quasi-real photons and therefore the experimental data on photodisintegration 
would be useless to us.  We would then have to resort to a full field 
theoretic calculation of the nuclear disintegration.  Such theory is both 
complicated and incomplete.  Therefore, this path is not very helpful.

Instead we will begin by transforming our magnetic field to the rest frame of 
the nucleus.  In this frame, the field looks very much like a
collection of real photons (hence the term quasi-real).  We express the field
as a collection of quasi-real photons and
 use the known experimental data on photodisintegration to calculate the 
likelihood that the nucleus will disintegrate in the magnetic field.

We begin with a static magnetic field $\bf{B} \it{(x,y,z)}$.  For simplicity 
we will Lorentz boost this in the x direction, so that the argument becomes
$x \to \gamma (x+vt)$, $y \to y$, $z \to z$.  The transformations of the
fields are also simple and can be found in any standard text such as 
\cite{PDG}.
\begin{eqnarray}
{\bf E'_\perp}&=& {\gamma}
\left ({\bf E_\perp} + \frac {\bf v} {c} \times {\bf
  B_\perp} \right ) \\ 
\bf{E'_\parallel}&=& \bf{E_\parallel} \\
{\bf B'_\perp}&=& {\gamma} 
\left ({\bf B_\perp} - \frac {\bf v} {c} \times {\bf
  E_\perp} \right ) \\
\bf{B'_\parallel}&=& \bf{B_\parallel} 
\end{eqnarray}  

We now substitute this expression into our original form for the magnetic
field and find
\begin{eqnarray}
{\bf B'} & = & \gamma {\bf B_\perp} + {\bf B_\parallel} \\
{\bf E'} & = & \gamma \frac{\bf v}{c} \times {\bf B}.
\label{eqn:restEM}
\end{eqnarray}

Here we note two important aspects of this electromagnetic field.  
First, the parallel component of the magnetic field is irrelevant.  It has no 
factor of $\gamma$ enhancement, and it has no partner electric field and 
therefore has no effect on a particle at rest.  This is expected as 
magnetic fields along the direction of a charged particle's motion generally 
have little or no effect.

Next, and more importantly, we notice that the electric field and the 
perpendicular component of the magnetic field are perpendicular to each
other and, for ultrarelativistic particles ($v \approx c$), approximately
the same magnitude.  Therefore, they can be viewed as a superposition of
quasi-real photons.  This is the subject of our next section.

\section {The Power Spectrum}

We now proceed to calculating the power and number spectrum of the
quasi-real photons.  Using eq. (15.51) of Ref.~\cite{Jackson}, we find that  
\begin{equation}
\frac {dI}{d\omega} = \frac {c} {2 \pi} 
 | {{\bf{E}} {(\omega)}}  |^2.
\end{equation}
Where $\bf{E} \it{(\omega)}$ is the Fourier transform of the electric field 
defined by
\begin{equation}
\bf{E} \it{(\omega)} = \frac {1} {\sqrt{2 \pi}} \int_{-\infty}^{\infty}
	{\bf{E}}(t) e^{i \omega t} {dt}.
\end{equation}
Substituting in equation \ref{eqn:restEM} for ${\bf{E}}(t)$ and changing 
variables we find
\begin{eqnarray}
{\bf{E'}}(\omega,x,y,z)& =& \int_{-\infty}^{\infty} {\bf{\beta \times B}}
	(\gamma(x+\beta c t),y,z) e^{i \omega t} {dt} \nonumber \\ & = & 
	\frac {1} {\beta c \sqrt{2 \pi}}  e^{i \frac {\omega x} {\beta c}}
	\int_{-\infty}^{\infty} {\bf{\beta \times B}}(x',y,z)
	 e^{i \frac {\omega x'} {\gamma \beta c}} {dx'}.
\end{eqnarray}
Where ${\bf{\beta}} = \frac {\bf {v}} {c} $.

This equation was derived with the assumption that the particle follows a
straight path through the magnetic field, but it is easy to generalize it.
The integral along $x$ will simply become a path integral along the
classical trajectory of the particle ${\bf{r}}(\ell)$, where $\ell$ is the
path length.  Then, droping factors of $\beta$ and the irrelevant phase
factor in front of the integral, we get
\begin{equation}
{\bf{E}}(\omega) = \frac {1} {c \sqrt{2 \pi}} \int_{\rm {path}}
	{\bf \beta \times B(r({\it {\ell}}))}
	e^{i \frac {\omega} {\gamma c} \ell} {d \ell}.
\label{eqn:Efourier}
\end{equation} 

\section{Photon flux and disintegration probability}

It is now straightforward to calculate the disintegration probability for the 
nucleus.  First, we relate the intensity to the flux of photons using
$$\frac {dI} {d\omega} = \hbar \omega \frac {dn} {d\omega}.$$ 

Theerefore, 
$$\frac {dn} {d\omega} = \frac {1} {\hbar \omega} \frac {c} {2 \pi}
	|{\bf{E}}(\omega)|^2.$$  
Next we multiply this flux by the experimentally determined cross
section for photodisintegration of the nucleus to find the differential
decay probability.
  
\begin{equation}
dP_{\rm{decay}} = 
	{\sigma}_{\gamma} (\omega) dn(\omega) =
	{\sigma}_{\gamma} (\omega)
	\frac {1} {\hbar \omega} \frac {c} {2 \pi} |{\bf{E}}(\omega)|^2 
	d{\omega}.
\label{eqn:Pdecay1}
\end{equation}

One change should be made at this point.  So far we have neglected the 
possibility that the particle has already decayed.  Thus, if we integrate 
equation 
(\ref {eqn:Pdecay1}) we may get answers greater than 1, which is impossible.
To rectify this we need to multiply the right hand side of the equation by
the probability that the particle has not disintegrated.  Thus,  equation 
(\ref {eqn:Pdecay1}) would be better written as
$$
dP_{\rm{decay}} = -dP_{\rm{survive}}=
	P_{\rm{survive}}{\sigma}_{\gamma} (\omega) dn(\omega).
$$
This can then be integrated to find that in fact the correct relation is
\begin{equation}
\ln{P_{\rm{survive}}}=
	-\int_{0}^{\infty} {\sigma}_{\gamma} (\omega) dn(\omega) =
	-\int_{0}^{\infty} {\sigma}_{\gamma} (\omega)
	\frac {1} {\hbar \omega} \frac {c} {2 \pi} |{\bf{E}}(\omega)|^2 
	d{\omega}.
\label{eqn:Pdecay2}
\end{equation}

\section {Various field configurations}
	
We now examine a variety of field configurations to see how
${\bf{E}}(\omega)$ depends on ${\bf{B}}(t)$.  We define 
$k=\frac {\omega} {\gamma c}$. 

\subsection{Gausian B-field} 

Let us suppose that the field is of the form ${\bf {B}}=\hat{j}B_0 \exp 
\left \{-\frac {x^2} {a^2} \right \}$ 
and the nucleus is traveling at ultrareletivistic speed 
in the $x$ direction.  From equation
(\ref{eqn:Efourier}) we find that
\begin{equation}
E(\omega)=\hat{k} \frac {B_0} {c \sqrt{2}} 
a \exp \left \{-\frac {\omega^2 a^2} {4 \gamma^2 c^2} \right \}
\end{equation}
We note that for sufficiently small values of $a$ this is relatively flat,
but that, for large values of $a$, $E(\omega)$ falls off very quickly with
increasing $\omega$.  However, for large enough values of gamma even very
large values of $\omega$ may have noticable contributions.

\subsection {Constant $B$-field with various cutoffs}     
	Next we consider a $B$-field that is relatively constant over a 
large region of size $a$ and which falls to $0$ over a much smaller region of 
size $b$.  The method we will use is to convolve a constant magnetic field 
with some approximation to a delta funtion, thus introducing a smoother cutoff.
This provides a nice way to compare different kinds of cutoffs on the same 
distance scale.  Also the Fourier transforms are then fairly easy to perform 
because the Fourier transform of a convolution is just the product of the 
Fourier transforms.  
For the rest of this subsection we will leave off the factor of 
$1/{c\sqrt{2\pi}}$ in equation (\ref{eqn:Efourier}).
\subsubsection{Constant Magnetic Field}
We'll start with a constant magnetic field:
$${\bf{B}_0}(x)=\left\{ \begin{array} {cc} {{\bf B}_0} & {x \in (-a/2, a/2)} \\
		{0} & {\rm otherwise.} \end{array} \right.$$
The Fourier transform of this is simply
$${\bf{B}}(k)=\frac {{\bf B}_0} {k} \, 
2 \sin \left ({\frac {ka}{2}}\right ).$$
If $ka\ll 2 \pi$ then this is nearly constant, however I expect that is 
unlikely.  In the opposite limit $ka \gg 2 \pi$ the oscilations in the sine 
function are enough that we can just replace it by it's root mean square value
of $\sqrt {2}$.

It is worth noting that for this configuration the $B$-field itself is 
discontinuous and the field falls off as $1/k$ for large values of $k$.

\subsubsection{Square pulse $\delta$-function}
From now on we will modify $\bf {B}$ by convolving it with some function 
$g(x,b)$ that has a characteristic width of $b$ and becomes a 
$\delta$-function in the limit that $b \to 0$.  (Basically this is just 
requiring that the area be normalized to 1.)  If $g$ were a true $\delta$-
function this would just give the same form we found before.  To state our 
assumptions mathematically
$${\bf{B}}(x)=\int_{-\infty}^{\infty}g(x-x'){\bf{B_0}}(x') dx'$$
$$\int_{-\infty}^{\infty}g(x-x')=1.$$

As our first approximate $\delta$-function we will use a square pulse.
$$g(x)=\left\{ \begin{array} {cc} {\frac {1} {b}} & {x \in (-b/2, b/2)} \\
		{0} & {\rm otherwise.} \end{array} \right.$$
The Fourier transform is simply $\tilde{g}(k)=
	\frac {\sin{(kb/2)}}{(kb/2)} $ so that the Fourier 
transform of the B-field is
\begin{equation}
{\bf{B}}(k)=\frac {{\bf B}_0} {k} 2 \sin {\frac {ka} {2}}
	\frac {\sin{(kb/2)}}{(kb/2)}.
\end{equation}

Note that the small $k$ $(k\ll1/b)$ behavior is essentially unchanged, but
that for large values of $k$ $(k\gg1/b)$ the field now falls off like
$1/k^2$.  Also note that whereas before the B-field itself was
discontinuous, now only its first derivative is discontinuous.
 
\subsubsection{Triangle pulse}

The next step is to consider a continuous approximate 
$\delta$-function with a
 discontinuous first derivative.  This will make the B-field continuous
 through the first derivative.  We find the following results.
$$g(x)=\left\{ \begin{array} {cc} {\frac {1} {b} (1+\frac {x} {b})}
					& {x \in (-b, 0)} \\
		{\frac {1} {b} (1-\frac {x} {b})}
					& {x \in (0, b)} \\
		{0} & {\rm otherwise.} \end{array} \right.$$
$$\tilde{g}(k)=\frac {2(1-\cos{kb})}{k^2b^2}=
	\left (\frac {\sin{(kb/2)}} {(kb/2)}\right )^2 $$
\begin{equation}
{\bf{B}}(k)=\frac {{\bf B}_0} {k} 2 \sin {\frac {ka} {2}}
	 \left ( \frac {\sin{(kb/2)}} {(kb/2)} \right )^2
\end{equation}
Again the behavior for small $kb$ is relatively stable, while for large $kb$
the function falls off as $1/k^3$

\subsubsection{Parabolic pulse}

Now we make the approximate $\delta$-function from
parabolas joined together in such a way that the function and first
derivative are both continuous.  Thus the B-field will have continuous 1st 
and 2nd derivatives.  Thus,
$$ g(x)=\left\{ \begin{array} {cc} {\frac {1} {b^3} (\frac {3} {4} b^2 -x^2)}
					& {x \in (-b/2, b/2)} \\
		{\frac {1} {2b^3} (x-\frac {3} {2} b)^2}
					& {x \in (b/2, 3b/2)} \\
		{\frac {1} {2b^3} (x+\frac {3} {2} b)^2}
					& {x \in (-3b/2, -b/2)} \\

		{0} & {\rm otherwise.} \end{array} \right.$$
$$\tilde{g}(k)= \left ( \frac {\sin{(kb/2)}} {kb/2} \right )^3$$
\begin{equation}
{\bf{B}}(k)=\frac {{\bf B}_0} {k} 2 \sin {\frac {ka} {2}}
	\left( \frac {\sin{(kb/2)}} {kb/2} \right )^3.
\end{equation}

Clearly, the observed pattern is continuing. I have chosen a very particular 
form for the parabolic pulse.  It is the pulse derived by convoluting the 
square and triangular pulses, and thus gives the very simple form listed.  
However, other parabolic pulses of width $b$ and unit area share the 
property that the Fourier transform goes smoothly to $1$ for small $k$ 
and falls off as 
$k^{-4}$ for large $k$. I am uncertain how to prove this pattern
in general, and the specifics are getting increasingly more difficult so I 
will not attempt smoother cutoffs of this form.

\subsubsection{Gaussian Pulse}
	It is; however, simple enough to explore the behavior of a 
particular 
perfectly smooth cutoff by using a gausian pulse as our approximate
 $\delta$-function.  This will yield a magnetic field that falls off quickly,
but with no discontinuous derivatives.  We expect to find that ${\bf B}(k)$ 
falls 
off more quickly than any power of $k$.  This is exactly what we find. 
$$g(x)= \frac {1} {b {\sqrt{\pi}}} \exp {-\frac {x^2} {b^2}}$$
$$\tilde{g}(k)=\exp{-\frac {k^2 b^2} {4}}$$
\begin{equation}
{\bf{B}}(k)=\frac {{\bf B}_0} {k} 2 \sin {\frac {ka} {2}}
		\exp{-\frac {k^2 b^2} {4}}
\end{equation}

\subsubsection{Summary and Discussion}

Table \ref{table:B(k)} summarises ${\bf B}(k)$ for the various field
configurations discussed in this subsection.  The last collumn simplifies
the form of ${\bf B}(k)$ by substituting a root mean square value for the
oscillatory part.

We should be clear at this point about what is meant by a discontinuity.
Certainly in any real situation we would expect the field and all of its
derivatives to be smooth down to the atomic scale.  However, from the point
of view of the Fourier transform, any change that occurs over a scale that
is small compared to the wavelength, $\lambda = 2 \pi / k$, will have the
same effect as a discontinuity.  This can be seen in table
\ref{table:B(k)}.  Each of the forms reduces to the case where the field
itself is discontinuous if the smoothing takes place over a distance that
is small compared to the wavelength.  In other words if $kb \ll 1$ or
equivalently $b \ll \lambda / {2 \pi}$ then the smoothing does not matter
and ${\bf B}(k)={\frac {{\bf B}_0} {k} 2 \sin {\frac {ka} {2}}}$.

By virtue of the Fourier transform, the more severe the discontinuity in 
${\bf B}(k)$, the slower the
fall-off in ${\bf B}(k)$.  If the first discontinuity in ${\bf B}$ or
it's derivatives appears at order ${\bf B}^{(i)}$ or the $i$th derivative
of ${\bf B}$, then for large $k$ the ${\bf B}(k)$ appears to fall off as 
$k^{-(i+1)}$.

\begin{table}[]
\caption{{\bf B} (k)}
\begin {tabular}{|c|c|c|c|} \hline
Smoothing Pulse & First Discontinuity & {\bf B}(k) & simplified {\bf B}(k) 
	\\ \hline
True $\delta$-function & {${\bf B}$} & 
  ${\frac {{\bf B}_0} {k} 2 \sin {\frac {ka} {2}}}$ & 
{${\frac {\sqrt{2}{\bf B}_0} {k}}$}
   \\ \hline
Square Pulse & ${\bf B}'$ & $\frac {{\bf B}_0} {k} 2 \sin {\frac {ka} {2}}
	\frac {\sin{( {kb}/ {2})}}{( {kb}/ {2})}$  & 
	{ $\frac {2 {\bf B}_0} {k^2 b} $} \\ \hline
Triangle Pulse & ${\bf B}''$ & $\frac {{\bf B}_0} {k} 2 \sin {\frac {ka} {2}}
	\left(\frac {\sin{( {kb}/ {2})}} {( {kb}/ {2})}\right )^2$ 
	& $\frac {2 \sqrt{3} {\bf B}_0} {k^3 b^2}$ 
	\\ \hline
Parabola Pulse & ${\bf B}'''$ &$ \frac {{\bf B}_0} {k} 2 \sin {\frac {ka} {2}}
	\left( \frac {\sin{(kb/2)}} {(kb/2)} \right )^3  $ & 
	$\frac {2 \sqrt{10} {\bf B}_0} {k^4 b^3} $\\ \hline
Gaussian Pulse & None & $\frac {{\bf B}_0} {k} 2 \sin {\frac {ka} {2}}
		\exp{-\frac {k^2 b^2} {4}}$ & 
		$\frac {\sqrt{2} {\bf B}_0} {k}
		\exp{-\frac {k^2 b^2} {4}}$ \\ \hline
\end{tabular}
\label{table:B(k)}
\end{table}

	The pattern that we see here appears to be a general property of
the Fourier transform that we are taking.  The Fourier transform is most
sensitive to structure (changes in magnetic field, discontinuities in the
field or its derivatives, etc.) that are on the same scale or smaller than
the wavelength.  The smoother the field is on this length scale, the more
heavily surpressed the Fourier transform is.  This means that in order to
see a significant effect on nuclei, the magnetic field will need to have
significant structure on the scale $\gamma \lambda_{dis}$ where
$\lambda_{dis}$ is a typical wavelength for photo- disintegration.  As we
shall see later, for the highest known cosmic-ray energies, this requires
structure on the order of a few milimeters.  However, the length scale
increases with increasing cosmic ray energy.  Also, repeating variations in
magnetic field, which would result in a smaller average field for other
measurements, would increase the size of ${\bf B}(k)$

\subsection{Dipole Field}

A dipole field can easily be represented by a scalar potential.  This is a
bit unconventional for a magnetic field, but it can be done, so long as
there are no currents.  (This way $\nabla \times {\bf B} = 0$.)  The field
${\bf B}$ is then simply ${\bf B} = \nabla \phi^*$.  We can now do the
Fourier transform on the single potential rather than on the three
components of the field.  To justify this, one can easily show that if
$$B(x,y,z)=\nabla \phi^*(x,y,z),$$ then
$$B_{x,y}(x,y,k)=\partial_{x,y} \phi^*(x,y,k)$$
$$B_{z}(x,y,k)=ik \phi^*(x,y,k).$$
The last term will not affect our calculation because it has no corresponding 
${\bf E}$-field. 

For a dipole field exterior to all currents generating the field,
the potential is
\begin{equation}
\phi^*=\frac {\vec {\mu} \cdot \vec{r}} {r^3},
\end{equation}
or in terms of components
$$\phi^*=\frac {\mu_x x + \mu_y y + \mu_z z} {[{z^2 + \rho^2}]^{\frac 3 2}}.$$
Where I have used $\rho=\sqrt{x^2+y^2}$ for simplicity when appropriate.
The Fourier transform of this configuration is

\begin{equation}
\phi^*(x,y,k)=2 \left [(\mu_x x + \mu_y y)\frac k \rho  K_1(k\rho)
	-i \mu_z k K_0(k \rho) \right ].
\label{previous_expression} 
\end{equation}
Where $K_i$ is a modified bessel function.

For $k\rho \gg 1$, the following  assymptotic expansion is valid: 
$$K_i(z)=\sqrt{\frac {\pi} {2z}} e^{-z} \left \{ 1 + \frac {4 i^2-1} {8z} +
	\ldots \right \}$$  
Therefore, one can simplify eq. (\ref{previous_expression}):  
\begin{equation}
\phi^*(x,y,k)=\left[\frac {\vec{\mu_\perp}\cdot\vec{\rho}} {\rho} - i\mu_z
	\right ]
	\sqrt{\frac {2 \pi k} {\rho}} e^{-k \rho}
\end{equation}
We can now take the gradiant of this (in the $x,y$-plane) to find ${\bf{B}}$.
\begin{equation}
{\bf B}(\rho, \theta, k)=\sqrt{\frac {2 \pi k} {\rho}} e^{-k \rho} \Biggl\{
\hat{\rho} \left[k(i \mu_z-\mu_\perp \cos{\theta})+ \frac {1} {2 \rho}(i
\mu_z-3\mu_\perp \cos{\theta}) \right ]+\frac {\vec{\mu_\perp}} {\rho}
\Biggr\}.
\end{equation}
It is worth noting that the leading behavior in $\rho$ for $k\gg1/ \rho$ is
\begin{equation}
{\bf B}(k) =\hat{\rho} k (i \mu_z-\mu_\perp \cos{\theta}) \sqrt{\frac {2
\pi k} {\rho}}e^{-k\rho}.
\end{equation}

\section{Required ${\bf B}(k)$ }
We now estimate the required $B$-field to get significant
photodisintegration due to the static fields, for the various field
configurations that we have described.  We use the approximate 
cross sections from Ref.~\cite{cross-sections}. The cross sections are
given in units of the classical dipole sum rule or $\sum_d=59.8 \frac {NZ}
{A}$~MeV-mb, and for almost all of the species listed, the integrated cross
section is about $1 \sum_d$ for single nucleon emission and an additional
$(.1-.2) \sum_d$ for double nucleon emission.  The cross sections generally
reach a maximum around $20$~MeV with a width of $8-12$~MeV.  Clearly, the
behavior is similar for all of these nuclei.  Therefore, we will simply
study an extremely simplified case.  We will take $\sum_d=60 \frac {NZ}
{A}$ and use $N$, $Z$, and $A$ for iron.  We will take the integrated cross
section to be $1 \sum_d$ centered at $20$~MeV with a width of $8$~MeV.  We
will ignore any variation in ${\bf B}(k)$ over this range and simply
substitute the value at $20 MeV$.  All of these approximations should be
correct to within $20\%$ for iron and will not result in any qualitative
changes in the result.

Using this data I calculated the magnetic field necessary to get a decay
probability of $1-e^{-1}$ for an iron nucleus with a gamma factor of
$10^{10}$ or an energy of about $5 \times 10^{20} eV$.  The results can be
found in table \ref{table:req-B}.  These results can be modified for other
values of gamma by multiplying $k$ times $\frac {10^{10}} {\gamma}$.  We
find that $k$ is approximately $100/$cm.  (Incidentally, this is close to
the shorter wavelength part of the CMB, which also starts to be important
at these energies. \cite{PDG2}) Entering this in equation (\ref{eqn:Pdecay2})
and restoring the $1/(c\sqrt{2\pi})$ we find that the necessary
$B(k)$ is
$${\bf B}(k)k = 1.7 \times 10^7 {\rm gauss}.$$ Substituting this into our
results from table \ref{table:B(k)} we can make estimates for the required
magnitude of the magnetic field for each of the field configurations
listed.  (Note: Discontinuity refers to the first derivative that changes
abruptly on the scale of $1/k$.)
\begin{table}
\caption{Required B-field for photodisintegration}
\label{table:req-B}
\begin {tabular}{|c|c|c|} \hline
Discontinuity & Required B field & Value \\ \hline
{$B$} & {$\frac {B(k) k} {\sqrt{2}}$} & {$1.2 \times 10^7 $} gauss \\ \hline
$B'$ & $\frac {B(k) k} {2} {kb}$ & $8.5 \times 10^{13} 
	\frac {\rm gauss} {\rm km} b$ \\ \hline
$B''$ & $\frac {B(k) k} {2 \sqrt{3}} {(kb)^2}$ & $4.9 \times 10^{20} 
			\frac {\rm gauss} {{\rm km}^2} b^2$ \\ \hline
$B'''$ & $\frac {B(k) k} {4 \sqrt{10}} {(kb)^3}$ & $2.6 \times 10^{27} 
			\frac {\rm gauss} {{\rm km}^3} b^3$ \\ \hline
None & $\frac {B(k) k} {\sqrt{2}} \exp{(\frac {(kb)^2} {4})}$ &
	$1.2 \times 10^7 \exp {(\frac {2.5 \times 10^{13} b^2} {\rm km^2})}$
		{\rm gauss}\\ \hline
\end{tabular}
\end{table}

It is difficult to imagine objects which would contain fields of this
magnitude. 
Since one would expect the magnetic field strengths to be
continuous and to fall off over scales of at least a few km, we need
magnetic fields of at least a $10^{14}$ gauss.  For smoother decreases,
larger objects, or motion that is not perpendicular to the magnetic field
the prospects get worse.  However, for higher energies or magnetic fields
with repeated large variations the effect may become noticable.

\section{Accelerating Particles}
All of the above treatment assumes that the particles are moving at a
constant speed through the magnetic field, and that photodisintegration
is caused by changes in the magnetic field.  It is also possible that
a particle is accelerated in the presence of a magnetic field, and that
the changes in its perceived magnetic field due to this acceleration 
could cause photodisintegration.  We discuss this possibility here.

If a particle feels a continuous proper acceleration $a$, (or 
equivalently, if it 
is pushed with a constant exterior force of $F=m_0 a$) then the 
relation between $\gamma$, $x$, and the proper time $\tau$ is as 
follows.

$$\gamma(\tau) = \cosh{(a\tau +b)}$$
$$x(\tau) = \frac {1} {a} \cosh{(a\tau + b)}+C$$
Here $b$ and $C$ are integration constants determined by the 
innitial conditions, and we have set the speed of light equal to $1$.

Hyperbolic functions can be difficult to work with, however, as long
as $\cosh{(a\tau + b)} \gg 1$ we can make the substitution
$\cosh{(a\tau + b)} \approx \frac {1} {2} \exp{(a\tau+b)}.$  Then
the above equations simplify to
\begin{equation}
\gamma(\tau)=\gamma_0 \exp{(a \tau)}
\end{equation}
\begin{equation}
x(\tau)=\frac {\gamma_0} {a} \bigl[\exp{(a\tau)}-1] + x_0.
\label{eqn:x-accel}
\end{equation}

We would like to follow the same procedure as in sections 2 and 3
to find a power spectrum for the effective electromagnetic field that
the particle sees.  Most of this except for the Fourier transform will
proceed as before.  However in finding 
$${\bf B}(\omega) = \int_{-\infty}^{\infty} {\bf B}(\tau) 
			e^{i \omega \tau} d\tau $$
a number of changes are in order.  First, during acceleration the 
relationship between $x$ and $\tau$ is now described by equation
(\ref{eqn:x-accel}).  Second, while it is reasonable to approximate a 
particle as always moving at a constant velocity, it is not 
reasonable to approximate it as always accelerating.  Therefore, we
must consider the particle's motion before, during, and after the 
acceleration.  It may be tempting to simply assume (correctly) 
that most of the contribution to the Fourier transform will come from
the acceleration, and therefore ignore the incoming and outgoing motion.
However, to do so would introduce spurious 
contributions as we shall see.  We will therefore assume that the 
particle comes in from $-\infty$ to point $x_0$ with $\gamma_0$, 
that it accelerates between $x_0$ and $x_1$ from $\gamma_0$ to 
$\gamma_1$, and that it finally leaves to $\infty$ with $\gamma_1$.
The correct form for the Fourier transform, after changing 
integration variables is then
\begin{eqnarray}
{\bf B} (\omega) &=& \int_{-\infty}^{x_0} {\bf B}(x) 
		e^{i \frac {\omega} {\gamma_0} (x-x_0)} dx \\
		&+&\int_{x_0}^{x_1} {\bf B}(x)
   	[{\frac {a} {\gamma_0}} (x-\tilde{x})]
		^{i \frac {\omega} {a}} dx \\
	&+&\bigl( {\frac {\gamma_1} {\gamma_0}} \bigr)^
				{i \frac {\omega} {a}}
	\int_{x_0}^{\infty} {\bf B}(x) 
		e^{i \frac {\omega} {\gamma_0} (x-x_1)} dx.
\end{eqnarray}
Where $\tilde{x}=x_0-\frac {\gamma_0} {a} = x_1-\frac {\gamma_1} {a}$.

If we assume that the magnetic field is constant except that it falls
off slowly at $\pm \infty$ so that we can ignore the integration limits
there, we find that 
\begin{equation}
{\bf B} (\omega) = \Biggl\{ \left( \frac {\gamma_1} {\gamma_0} \right )
			^{i \frac {\omega} {a}}
		\gamma_1 - \gamma_0 \Biggr\}
	\biggl[\frac {1} {i\omega+a} - \frac {1} {i\omega} \biggr]
	=\Biggl\{ \left( \frac {\gamma_1} {\gamma_0} \right )
			^{i \frac {\omega} {a}} 
		\gamma_1 - \gamma_0 \Biggr\}
		\frac {a} {\omega^2-i a \omega}.
\end{equation}
Where the first term in $[\hspace{5pt}]$ comes from the acceleration
integral and the second term comes from the incoming and outgoing
integrals.  Notice that as mentioned before, the overall behavior in
$\omega$ is softer than in either term alone.  If we had integrated only
over the acceleration region we would have essentially introduced a
fictitious abrupt cutoff in the magnetic field.  The incoming and outgoing
motion cancels this effect.  We can see that if the acceleration is much
larger than $\omega$ the falloff with $\omega$ is fairly slow, and the
problem becomes similar to an abrupt falloff in the B-field.  As can be
seen in table \ref{table:req-B} this may be an achievable magnetic field in
some extreme environments.  However, such high accelerations require that
the particle's energy change by nine orders of magnitude in a distance of a
few microns.  Thus, we expect much softer behavior, which would require
much larger magnetic fields to achieve photodisintegration.

\section{Conclusion}
It apears unlikely that static magnetic fields would cause significant 
photo-disintegration of nuclei, unless the fields are rapidly changing 
or the particles are accelerating, on scales of a few microns to a few
millimeters.  Not coincidentally, this is the same scale as the shorter
wavelengths of the CMB~\cite{PDG2} which cause attenuation of such high energy 
cosmic rays as they traverse the universe.

While this suggests that static magnetic fields are unlike to cause large
energy losses in cosmic ray nuclei, the effect may become more significant
at higher energies, or if the magnetic fields experience repeated
variations.

\section{Acknowledgements}

I'd like to thank A. Kusenko for helpful discussions and editing.
This work is supported in part by the DOE grant DE-FG03-91ER40662.

\end {document}